# Impacts of DEM Type and Resolution on Deep Learning-Based Flood Inundation Mapping

Mohammad Fereshtehpour[1]✉ , Mostafa Esmaeilzadeh[2], Reza Saleh Alipour[3]; Steven J. Burian[3]

[1]Department of Civil and Environmental Engineering, University of Western Ontario, London, Canada

[2]Department of Water and Environmental Engineering, Faculty of Civil Engineering, Shahrood University of Technology, Shahrood, Iran

[3]Department of Civil, Construction, and Environmental Engineering, University of Alabama, Tuscaloosa, United States

✉mferesht@uwo.ca

## Abstract

The increasing availability of hydrological and physiographic spatiotemporal data has boosted machine learning's role in rapid flood mapping. Yet, data scarcity, especially high-resolution DEMs, challenges regions with limited access. This paper examines how DEM type and resolution affect flood prediction accuracy, utilizing a cutting-edge deep learning (DL) method called 1D convolutional neural network (CNN). It utilizes synthetic hydrographs as training input and water depth data obtained from LISFLOOD-FP, a 2D hydrodynamic model, as target data. This study investigates digital surface models (DSMs) and digital terrain models (DTMs) derived from a 1 m LIDAR-based DTM, with resolutions from 15 to 30 m. The methodology is applied and assessed in an established benchmark, the city of Carlisle, UK. The models' performance is then evaluated and compared against an observed flood event using RMSE, Bias, and Fit indices. Leveraging the insights gained from this region, the paper discusses the applicability of the methodology to address the challenges encountered in a data-scarce flood-prone region, exemplified by Pakistan. Results indicated that utilizing a 30 m DTM outperformed a 30 m DSM in terms of flood depth prediction accuracy by about 21% during the flood peak stage, highlighting the superior performance of DTM at lower resolutions. Increasing the resolution of DTM to 15 m resulted in a minimum 50% increase in RMSE and a 20% increase in fit index across all flood stages. The findings emphasize that while a coarser resolution DEM may impact the accuracy of machine learning models, it remains a viable option for rapid flood prediction. However, even a slight improvement in data resolution in data-scarce regions would provide significant added value, ultimately enhancing flood risk management.

*Keywords:* Flood inundation, DEM resolution, Convolutional Neural Network, Rapid prediction, Machine learning, Artificial intelligence.

## Published Version

# 1. Introduction

Natural disasters cause great damage to human societies every year. According to the World Meteorological Organization (WMO), floods are the third most impactful disaster in the world and have the highest number of deaths and injuries (Zahir et al., 2019). In addition, the impacts cascade to the buildings, transportation infrastructure, critical facilities, cultural heritage, environment, and economy, (Yu et al., 2013). Globally, the frequency of floods has increased by 40% in the last two decades (Khosravi et al., 2019). The impacts of floods may be mitigated by decision-makers with accurate and timely flood forecasting information (Chen et al., 2018).

Flood management can be divided into four main processes: forecasting (Mosavi et al., 2018), detection (Van Ackere et al., 2019), mapping (Manavalan, 2017), and risk assessment. With more hydrologic and physiographic data available in recent years, machine learning (ML) techniques have proven to be a reliable tool that helps improve flood forecasting models, post-flood mapping, and complex physical and dynamic modeling efforts (Mosavi et al., 2018; Muñoz et al., 2021). ML models are defined within the context of flood risk assessment to provide alternatives and complements to the historical disaster mathematical statistics method (HDMS), scenario simulation analysis (SSA), and multi-criteria decision analysis (MCDA) (Chen et al., 2021). HDMS assesses historical disasters reported by governments and officials and gives results that are generally consistent with reality (Xu et al., 2018; Wang et al., 2020; Chen et al., 2021). SSA typically requires 2D hydraulic/hydrodynamic models for high-sensitivity flood zone detection, implementing through software such as Mike (Banks et al., 2014; Kourtis et al., 2017; Jahandideh et al., 2020), Info Works ICM (Sameer and Rustum, 2017; Liu et al., 2021), and LISFLOOD-FP (Bates and De Roo, 2000; Wu et al., 2017; Grimaldi et al., 2019; Liu et al., 2019), among others. The MCDA provides a flexible flood risk assessment plan that has been implemented in many areas (Tang et al., 2018; Mishra and Sinha, 2020; Pham et al., 2021,2022). ML models assess floods by automatically learning past flood risk characteristics and predicting future conditions (Costache et al., 2019, 2021; Nevo et al., 2022). ML models are equally adaptable as MCDAs but yield more objective results. These models are often better suited for predicting the likelihood and severity of future floods based on historical data, while SSA is better suited for testing the effectiveness of different flood management strategies under different conditions.

The majority of SSA models solve a simplified version of the Navier-Stokes equation (NS), coupled with the conservation of mass equation, to formulate the motion of a fluid (Janna, 1993). Solving differential equations of conservation of mass and momentum for a large domain (national or continental scale) is tedious, time-consuming, and costly (Rahman et al., 2021). This makes the use of hydraulic models for large-scale flood simulations generally impractical, especially when using fine-resolution topographic data in urban areas and other heterogeneous locations (Woznicki et al., 2019), which requires substantial simplifications in the modeling. Such simplifications increase the uncertainty and inaccuracy of predicted results (Zarzar et al., 2018). As a result, running simulations in real-time, such as those needed for emergency response, using the hydraulic model is almost unfeasible (Maidment et al., 2016). Therefore, there is a need for quick, robust, and versatile approaches for large-scale, real-time flood modeling.

To mimic the complex mathematical expressions of physical processes of floods, during the past two decades, ML methods have contributed to the advancement of prediction systems providing better performance and cost-effective solutions (Mosavi et al., 2018). Such approaches are specifically useful when (a) the current models are not fully capable of capturing the physics in mathematical terms, (b) the computational cost is impractical, or (c) the available knowledge about the problems is limited (Hosseiny et al., 2020). This has made ML approaches powerful tools for assessing different aspects of flood risk assessment.

ML techniques have evolved through time (Bhattacharya et al., 2007), focusing on learning from current data and experiences to enhance understanding of real-world problems (Mitchell, 1997). Using new and advanced ML algorithms ultimately provides a model with greater accuracy (i.e., the overall correctness of a model's prediction) and sensitivity (i.e., the model's ability to correctly identify positive cases or instances of a particular class or condition) (Arora et al., 2021). Recently, new techniques and methods have been suggested based on ML (Kabir et al., 2020; Pham Quang and Tallam, 2022; Antwi-Agyakwa et al., 2023). Among all ML methods, ANNs are the most popular learning algorithms, known to be versatile and efficient in modeling complex flood processes with a high error tolerance and accurate approximation (Mosavi et al., 2018). In light of the difficulty of using ML approaches for precise and trustworthy flood forecasting, several neural networks such as Adaptive Neuro-Fuzzy Inference System (ANFIS), Multilayer Perceptron (MLP), Wavelet Neural Network (WNN), Ensemble Prediction System (EPS), Decision Tree (DT), Random Forest (RF), Classification And Regression Trees (CART) and Convolutional Neural Network (CNN) have been suggested as potential solutions for strengthening the reliability of warning systems without sacrificing accuracy (Khosravi et al., 2019; Arora et al., 2021; Nevo et al., 2022; Chen et al., 2022). When combined with image processing methods, the instruments available for flood control, both before and after a natural catastrophe become far more advanced (Munawar et al., 2021; Singha and Swain, 2022; AlAli and Alabady, 2022; Zabihi et al., 2023).

In recent years, several studies have been conducted to develop an ML-based flood risk analysis framework. Table 1 summarizes the most important and relevant studies, indicating the inputs and outputs of the ML model and the role of hydraulic models within the proposed framework.

The majority of these studies have touched on how various machine learning methods may be used in flood prediction and early warning systems (Noymanee et al., 2017; Chen et al., 2018; Mosavi et al., 2018; Woznicki et al., 2019; Park et al., 2020; Arora et al., 2021; Avand et al., 2022). Few research studies have used hydraulic models and ML, even though the combined models are substantially more reliable than the other cases (Liu et al., 2017; Jhong et al., 2017; Kabir et al., 2020). In the realm of ML techniques, shallow learning methods have received considerable attention in research such as K-Nearest Neighbors (KNN) (Park et al., 2020), Kernel Functions (KF) (Liu et al., 2017), Support Vector Machine (SVM) (Jhong et al., 2017; Chen et al., 2021), Support Vector Regression (SVR) (Jhong et al., 2017; Kabir et al., 2020), ANN (Mosavi et al., 2019; Shafizadeh-Moghadam et al., 2018), MLP (Chen et al., 2021; Avand et al., 2022, Karamouz et al., 2022), ANFIS (Arora et al., 2021), RF (Woznicki et al., 2019; Gudiyangada et al., 2020), and Decision Tree (DT) (Gharakhanlou and Perez, 2023). In contrast, the application of deep learning algorithms for flood prediction has not been thoroughly explored and warrants further investigation.

In flood management, DL algorithms can be valuable tools for analyzing flood data and predicting potential flood events. These algorithms can process large amounts of data and identify patterns that may not be apparent to human analysts and other techniques, leading to more accurate predictions and actionable insights (Pally and Samadi, 2022). Hydrological studies have benefited from different architectures of DL, such as recurrent neural networks (RNNs) (Wang et al., 2020), long short-term memory (LSTM) (Kratzert et al., 2019; Nevo et al., 2022; Adaryani et al., 2022; Foroumandi et al., 2023), and convolutional neural networks (CNNs) (Ni et al., 2020; Chang et al., 2022). Since CNN accelerates feature extraction and spatial analysis, it has garnered considerable interest in flood prediction (Kabir et al., 2020; Moy de Vitry et al., 2019; Wang et al., 2019; Kabir et al., 2020). In general, CNNs are counter-neural networks with alternating layers of convolution and subsampling and mainly trained in a supervised manner. CNN's superiority lies in its capacity for self-learning, whereby it can autonomously learn and organize features from vast datasets via multiple neuron layers (Munawar et al., 2021). Additionally, CNNs have exhibited exceptional aptitude in tasks such as image classification, segmentation, and feature extraction, as evident from various studies (Chang et al., 2014; Bhandare et al., 2016).

Recent studies have demonstrated the potential of CNN models in flood risk management. In a comparative analysis by Kabir et al. (2020), the CNN model outperformed the Support Vector Regression (SVR) method for predicting flood depth in the urban landscape of Carlisle, UK. Liu et al. (2021) used a 1D-CNN model to overcome the learning process challenges of traditional artificial neural network (ANN) models and accurately identified flood-prone areas. Pally and Samadi (2022) employed CNNs to classify, label, and weigh flood data from images captured by surveillance cameras and geographical information, enabling them to determine the depth, intensity, and risk of flooding in sensitive areas. These studies suggest that CNNs may be a promising tool for flood risk management, enabling more accurate predictions and identification of flood-prone areas.

In mapping the spatial extent of floods, the Digital Elevation Model (DEM), which provides information on the elevation and slope of the terrain, is proven to be a critical input (Saksena and Merwade, 2015; Karamouz and Fereshtehpour, 2019; Muthusamy et al., 2021; Avand et al., 2022). This data has been extensively used in the ML-based flood risk assessment models (Park and Lee., 2020; Muñoz et al., 2021; Chen et al., 2021; Saha et al., 2021; Lin et al., 2022; Gharakhanlou and Perez, 2023; Zeng and Bertsimas, 2023). However, in this context, there has been a scarcity of research that has specifically examined the impact of DEM spatial resolution and type on flood prediction. Recently, Saha et al. (2021) investigated the impact of different image and DEM data resolutions on flood sensitivity predictions using several shallow learning techniques, including ANN-multilayer perceptron (MLP), random forest (RF), bagging (B)-MLP, B-gaussian processes (B-GP), and B-SMOreg. They identified elevation, drainage density, and flow accumulation as the most significant factors affecting flood sensitivity. Avand et al. (2022) conducted a study to investigate the impact of spatial resolution of DEMs on the accuracy of flood probability prediction using three machine learning models, namely RF, ANN, and GLM. They highlighted the importance of careful selection of DEM spatial resolution based on specific modeling objectives and applications. While previous studies on flood prediction have shown the potential of shallow machine learning techniques, the use of deep learning methods has been identified as a promising approach for improving the accuracy of flood forecasts. However, there is a significant research gap regarding the impact of DEM resolution on the performance of deep learning models for flood prediction. This is especially important in areas where high-resolution DEM data is not readily available.

This paper aims to evaluate the influence of DEM type and resolution on the rapid prediction of fluvial flood inundation using a state-of-the-art 1-D CNN-based model. The city of Carlisle in the UK serves as the case study which is used extensively as a benchmark in flood risk research. The study encompasses two key objectives. Firstly, it aims to investigate how the type and resolution of DEMs affect flood depth and extent within the framework of deep learning methods. Secondly, it aims to underscore the critical importance of providing higher-resolution DEMs in regions with limited data availability, which are increasingly vital due to the escalating demand for machine learning techniques.

**Table 1.** Summary of the literature on flood prediction using Machine Learning

| Authors | Study Area | Inputs | ML technique | ML output |
|---|---|---|---|---|
| Liu et al. (2017) | Huai River, China | Historical hydrographs | KNN[1], KF[2] | Flood hydrograph |
| Jhong et al. (2017) | Chiayi, Taiwan | Flow depth, Land use, DEM | SVM[3], SVR[4] | Inundation map |
| Noymanee et al. (2017) | Pattani river, Thailand | Water level, Rainfall, Flash flood hydrograph, Riverbed data | MS Azure ML[5] | Water Level, Flood Peak |
| Chen et al. (2018) | Yangtze River, China | Flow data, Runoff | ELM[6], ELM-BSA[7], GRNN[8] | Stream Flow Rate |
| Mosavi et al. (2018) | ____ | River flows, River water level, Rainfall, Historical event hydrographs | ANN[9], MLP[10], ANFIS[11], WNN[12], SVM, DT[13], EPSs[14] | Inundation map |

---

[1] K-Nearest Neighbor (KNN)
[2] Kalman Filter (KF)
[3] Support Vector Machine (SVM)
[4] Support Vector Regression (SVR)
[5] Microsoft Azure Machine Learning (MS Azure ML)

[6] Extreme Learning Machine (ELM)
[7] Extreme Learning Machine-Backtracking Search Algorithm (ELM-BSA)
[8] General Regression Neural Network (GRNN)
[9] Artificial Neural Networks (ANN)

[10] Multilayer Perceptron (MLP)
[11] Adaptive Neuro-Fuzzy Inference System (ANFIS)
[12] Wavelet Neural Network (WNN)
[13] Decision Tree (DT)
[14] Ensemble Prediction Systems (EPSs)

**Table 1.** Continued.

| Authors | Study Area | Inputs | ML technique | ML output |
|---|---|---|---|---|
| Moghadam et al. (2018) | Haraz watershed, Iran | TWI[15], SPI[16], Land use, NVDI[17], Rainfall and Petrology, Slope degree, Curvature, Datasets DEM-20m[18], Distance to River | ANN, CART[19], FDA[20], GLM[21], GAM[22], BRT[23], MARS[24], MaxEnt[25], | Flood sensitivity map |
| Woznicki et al. (2019) | Floodplains, United States | FIRMs[26], Datasets DEM-30m[27] | RF | Inundation map |
| Kim et al. (2019) | Seoul, South Korea | Rainfall data, Overflow | NARX[28], SVNARX[29], SOFM[30] | Inundation map |
| Khosravi et al. (2019) | Ningdu, China | Flood Inventory Map, NDVI, Land use, Distance from river, Curvature, Altitude, TWI, SPI, Soil type, Slope, Rainfall | NBT[31], NB[32], SAW[33], TOPSIS[34], VIKOR[35] | Flood sensitivity map |

---

[15] Topographic Wetness Index (TWI)
[16] stream power index (SPI)
[17] Normalized Vegetation Difference Index (NDVI)
[18] Datasets Digital Elevation Model-20 m
[19] Classification And Regression Trees (CART)
[20] Flexible Discriminant Analysis (FDA)
[21] Generalized Linear Model (GLM)
[22] Generalized Additive Model (GAM)

[23] Boosted Regression Trees (BRT)
[24] Multivariate Adaptive Regression Splines (MARS)
[25] Maximum Entropy (MaxEnt)
[26] Flood Insurance Rate Maps (FIRMs)
[27] Datasets Digital Elevation Model-30m
[28] Nonlinear Auto-Regressive with Exogenous (NARX)
[29] Second Verification Algorithm of Nonlinear Auto-Regressive with Exogenous (SVNARX)

[30] Self-Organizing Feature Map (SOFM)
[31] Naïve Bayes Trees (NBT)
[32] Naïve Bayes (NB)
[33] Simple Additive Weighting (SAW)
[34] Technique for Order Preference by Similarity to Ideal Solution (TOPSIS)
[35] Vise kriterijumska optimizacijaik ompromisno Resenje (VIKOR)

**Table 1.** Continued.

| Authors | Study Area | Inputs | ML technique | ML output |
|---|---|---|---|---|
| Park et al. (2020) | Coastal areas, South Korea | Rainfall, Elevation, Slope, Tide data | KNN, RF, SVM | Risk probability map |
| Nachappa et al. (2020) | Salzburg, Austria | Flood Inventory Map, Distance to Drainage, Rainfall, Elevation, Slope, NDVI, SPI, TWI, Distance to Roads, Land Cover | RF, SVM | Flood sensitivity map |
| Kabir et al. (2020) | Carlisle, UK | Water Depth (from LISFLOOD-FP), Hydrographs | CNN, SVR | Inundation map |
| Chen et al. (2021) | Pearl River, China | DEM, Slope, Distance to River, Road Density, TWI, Curve Number | SVM, RF, GBDT[36], XGBoost[37], MLP, CNN | Flood risk map |
| Arora et al. (2021) | Middle Ganga Plain, India | Flood inventory data, Conditioning factors (CgFs) | ANFIS & ANFIS-GA (Genetic Algorithm), ANFIS-DE (Differential Evolution), ANFIS-PSO (Particle Swarm Optimization) | Flood hazard classification |

---

[36] Gradient Boosting Decision Tree (GBDT)          [37] eXtreme Gradient Boosting (EGBoost)

**Table 1.** Continued.

| Authors | Study Area | Inputs | ML technique | ML output |
|---|---|---|---|---|
| Avand et al. (2022) | Tajan watershed, Iran | Rainfall, Land use, Drainage density, Slope, Distance from the river, Altitude of the area | SOM, RBFNN[38], MLP | Flood susceptibility map |
| Nevo et al. (2022) | India and Bangladesh | Rainfall, Water level, DEM | LSTM[39], linear models, | Inundation map |
| Chen et al. (2022) | Xinyang, China | Longitude& latitude, Rainfall, Discharge | CNN, LSTM, Conv LSTM[40] | Streamflow rate |
| Chang et al. (2022) | Taipei, Taiwan | Storm-induced rainfall data, Rainfall pattern, Water depth | PCA[41], SOM, NARX | Inundation map |
| Gharakhanlou and Perez (2023) | Nicola & Fraser River, British Columbia | Lithology, Drainage density, Distance from rivers, Soil, Rainfall, Land cover, NDMI[42] | DT, RF, MLP-NN[43], AdaBoost[44], LR[45], SVM | Flood susceptibility map |

---

[38] Radial Basis Function Neural Network (RBFNN)
[39] long short-term memory (LSTM)
[40] convolution LSTM (Conv LSTM)

[41] Principal Component Analysis (PCA)
[42] Normalized Difference Moisture Index (NDMI)
[43] Multilayer Perceptron Neural Network (MLP-NN)

[44] Adaptive Boosting (AdaBoost)
[45] Logistic Regression (LR)

## 2. Methodology

To assess the performance of the CNN-based flood prediction model when using different types and resolutions of DEMs, a hydrological and topographic dataset is first generated. Using the observed and synthetic hydrographs and DEM dataset, the hydrodynamic model simulations are executed to derive the time series of water depth at each cell in the domain for the entire flood duration. The upstream synthetic hydrographs and the hydraulic simulation output (as the baseline rather than field observations) are then used as the input and target variables for training the CNN-based model, respectively. The trained ML model for each DEM will then be evaluated for the pseudo-observed inundation obtained from the higher-resolution DEM. Figure 1 shows the workflow of the present study. The following subsections describe the details of the models and parameters used in this study.

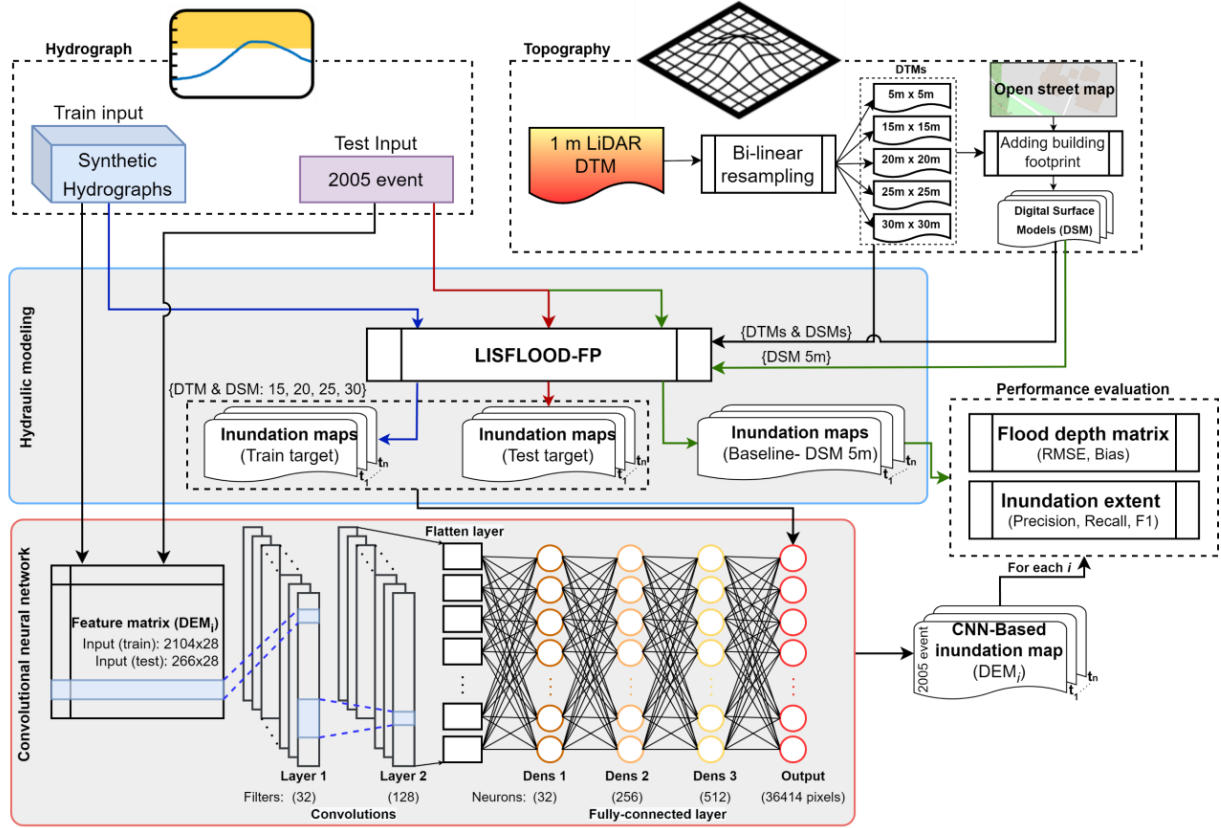

**Figure 1.** A schematic diagram explaining the workflow of the present study. The colors of the arrows represent the workflow from the input to its corresponding output. It is important to note that the number of outputs in the convolutional neural network corresponds to the number of pixels for each resolution. As an example, the displayed "364,141 pixels" is associated with a spatial resolution of 20 m.

## 2.1 Convolutional neural networks

A convolutional neural network (CNN) is a subcategory of ANNs that takes its design cues from the visual brain of living organisms (Hu et al., 2015; Kiranyaz et al., 2021). When compared to other ML methods, CNN has the benefits of broad application, parallel processing capability, and weight sharing, which means that global optimization training parameters are substantially reduced (Vedaldi and Lenc, 2015; Chen et al., 2018). CNNs are essentially multilayer feed-forward neural networks that can automatically extract useful characteristics from raw input (Zhang et al., 2018; Wang et al., 2020). The basic components of a CNN model are an input layer, one or more hidden layers, and an output layer; the hidden layers may include either convolutional or pooling layers (LeCun et al., 2015; Ghorbanzadeh et al., 2019). The convolutional layer, which is made up of several convolution kernels, pulls complex and useful characteristics from the original input repeatedly (Canziani et al., 2016; Mallat, 2016; Kiranyaz et al., 2021). After a convolutional layer, the pooling (sub-sample) layer is often used to decrease the dimensionality of feature maps using a down-sampling technique. This may help prevent overfitting and lower computing costs (Chen et al., 2016). The backpropagation (BP) algorithm is most often used to train CNNs in a supervised way (Kiranyaz et al., 2021). The gradient magnitude (or sensitivity) of each network parameter, such as the weights of the convolution and fully connected layers, is calculated during each iteration of the BP. The CNN parameters are then repeatedly updated using the parameter sensitivities until a predetermined stopping threshold is satisfied (Chauvin and Rumelhart, 2013; Ozcan et al., 2022).

It is known that classic deep CNNs are only intended to work with 2D data such as images and videos (Kiranyaz et al., 2021). A recent development is a modified form of 2D CNNs known as 1D Convolutional Neural Networks (1D CNNs) (Kiranyaz et al., 2015; Abdeljaber et al., 2018; Harbola and Coors, 2019; Yang et al., 2023; Huang et al., 2023). The cited studies have proven that when dealing with 1D signals, 1D CNNs are beneficial and hence preferred to their 2D equivalents. Reasons for this include the need for specialized hardware configuration for training 2D CNN and the high computational complexity of 2D convolution and its deep architecture with more than 10M parameters compared to that of 1D CNN (typically fewer than 10K parameters) (Kiranyaz et al., 2021).

In 1D convolutional neural networks, two types of layers have been proposed: 1) CNN-layers which perform 1D convolutions, activation functions, and sub-sampling (pooling), 2) Layers that are fully connected (dense) and similar to the layers of a standard multi-layer perceptron (MLP), which are referred to as "MLP-layers" (Hou et al., 2018; Ozcan et al., 2022). Each layer's filter (kernel) size, subsampling factor, pooling, and activation functions, as well as the total number of hidden CNN and MLP layers/neurons, are all crucial hyper-parameters that make up a 1D configuration (Ni et al., 2023).

This study employs the 1D-CNN model developed by Kabir et al. (2020) based on the Python programming language. With a 5 m DEM, the model performed well in validation, with an RMSE of 0.11 meters and a Nash-Sutcliffe efficiency coefficient of 0.86 (Kabir et al., 2020). Figure 1 depicts the main structure of the 1D-CNN model. The network comprises five hidden layers, two of which are convolutional and three of which are dense. The dense layers are fully connected and function similarly to an MLP network. The output layer comprises nodes equal to the number of cells in the simulation domain, while the input layer receives the upstream flow discharge data. Details of specific model parameters are discussed in the study area section.

## 2.2 Hydraulic modeling

LISFLOOD-FP is a raster-based model for simulating fluvial or coastal flood inundation that was first proposed by Bates and De Roo (2000). Thanks to its remarkable computational efficiency, the model has been extensively developed and tested effectively in several case studies across the world (e.g., Neal et al., 2011; Amarnath et al., 2015; Sosa et al., 2020). With its first release, LISFLOOD-FP used an explicit forward difference approach on a staggered grid in a 2D plane to solve the zero-inertial approximation of the Saint Venant equations (also known as the diffusion-wave approximation) (Bates and De Roo, 2000). In order to simulate floodplain hydrodynamics, a re-formulation of the LISFLOOD-FP model version 6.3.1 (Neal et al., 2012) is utilized, which solves the shallow water equations while simply ignoring the convective acceleration element (Bates et al., 2010). LISFLOOD-FP employs a sub-grid river channel representation for locations with insufficient channel information. Channels are 1D sub-grid scale features that may be wider or narrower than the floodplain grid. The development of floodplain inundation in two dimensions is modeled when the channel water depth surpasses the channel bank elevation and overflows to the overlying structured grid (O'Loughlin et al., 2020). In addition to the aforementioned features, the model's remarkable flexibility, coupled with its ability to effectively handle few input datasets and its straightforward input and output format, renders it highly compatible with various flood risk assessment frameworks. Moreover, as an open-source model, the model benefits from continuous development and improvement driven by a community of experts, ensuring that it stays up to date with the latest advancements in flood modeling. Notably, the recent implementation of new GPU-accelerated solvers has significantly expedited flood simulations, thereby enabling its applicability in urban and catchment scale modeling scenarios (Sharifian et al., 2023).

To demonstrate the performance of the ML-based predictive models, it is necessary to generate sufficient input (upstream boundary conditions, i.e., flow hydrographs) and output (water depth) data to train the ML models. For this purpose, multiple synthetic hydrographs with various peaks and durations are produced to represent flood scenarios of different magnitudes. The next important input data source for hydraulic models is the digital elevation model (DEM) (Dutta and Herath, 2001; Saksena and Merwade, 2015; Karamouz and Fereshtehpour, 2019). In this study, the DTM of various spatial resolutions is produced using the highest DEM resolution available using the bilinear interpolation approach. Then to produce the DSMs, the building footprints are added to the constructed DTMs. Given the importance of buildings in influencing flood routes in urban settings, the proposed DEM correction approach would concentrate on these critical urban terrain elements. Buildings are represented in the raw DEM by raising the topographic elevation to a specific height, thereby creating 'islands' that hinder flood water as they do in reality (Xing et al., 2022). Since we only included buildings, the current method could be sufficient. However, the accuracy of the building footprints used in the correction process can impact the quality of the corrected DEM. Moreover, the constant height assumption may not always hold true, as buildings and infrastructures can have varying heights and the terrain can be complex, thus, taking into account height adjustments would be more beneficial.

Considering DTMs and DSMs of different spatial resolutions, the LISFLOOD-FP is executed while employing sets of synthetic inflow hydrographs. The output of this process would be grid-based distribution of flood depths, which would serve as the target dataset for training the predictive model.

## 2.3 Predictive model

The CNN model is trained on each of the DEM datasets by utilizing the input (predictor) and output (target) variables from the synthetic flood scenarios. By transforming the synthetic hydrographs at each upstream location and the accompanying raster-based flood depths predicted by LISFLOOD-FP into matrices, both the input and the output variables can be defined. In this study, the primary inputs for predicting water depths are discharge values with $n_{ts}$ antecedent time-steps for each of the upstream locations $i$ and the associated observation time. As a result, there are a total of $(n_{tc} + 1)n_i$ input variables. The input feature matrix is constructed by vertically stacking the discharge with antecedent values and associated observation periods for all synthetic hydrographs. After transforming the consecutive water depth raster data into arrays, they are stacked vertically to form the target matrix. The input and target variables are then used to train the ML models, with each row of the input and output matrices being considered as a sample of the training dataset. It should be noted that the CNNs are trained on an Intel I5-8250U CPU with four cores. In the last step, the test datasets are fed into the CNN-based model that has already been pre-trained to predict water depths over the whole study area.

In order to prevent the CNN model from overfitting, several regularization techniques are used during the training phase, such as "early stopping", "batch normalization" and "dropout" (Kabir et al. 2020). "Early stopping" was implemented to prevent the model from learning irrelevant patterns or noise in the data (Prechelt, 2002). It halts the training process when the model's performance on a validation dataset no longer improves, avoiding overfitting and finding the optimal point for generalization improvement. "Batch normalization" was employed to stabilize the training process by normalizing the inputs of each layer in the neural network (Kim and Panda, 2021). This technique addresses issues related to internal covariate shift, leading to faster convergence and improved generalization by reducing reliance on specific parameter initializations. "Dropout" was incorporated as a regularization method to prevent co-adaptation among neurons. By randomly dropping out a fraction of neurons during each training iteration, this technique encourages the network to learn robust features and reduces the risk of overfitting by avoiding dependency on specific neurons (Srivastava et al., 2014). As such, the hyper-parameter settings were determined through a process of experimentation and fine-tuning, aiming to achieve the optimal balance between model performance and preventing overfitting. For the "early stopping" technique, the settings for patience (the number of epochs to wait before stopping) and the improvement criteria (min delta) were configured as 5 and 0.001, respectively. Additionally, a batch size of 10 and a dropout rate of 0.2 were chosen. The Adam optimizer was employed for training the network, with a learning rate set to 0.01. Moreover, as depicted in Figure 1, the convolutional layers in the network utilized filters of size 28 and 128. In the fully connected layers, the number of neurons were set to 32, 256, and 512.

## 2.4  Performance assessments

The outputs from the employed hydraulic model are directly compared with the ML predictions to assess the performance of the CNN-based flood depth prediction model in simulating the outcomes of a 2D hydraulic model. Two metrics—RMSE and Bias—are used in this study to determine the CNN model's efficacy in predicting flood depth, and three additional metrics—Recall, Precision, and $F_1$ score—are used to demonstrate how well the model performs in identifying the correct binary inundation status for each computational pixel. The root mean squared error (RMSE) (Boyle et al., 2000) and Bias are defined as

$$RMSE = \sqrt{\frac{1}{n}\sum_{i=1}^{n}(S_i - O_i)^2} \tag{1}$$

$$Bias = \frac{\sum_{i=1}^{n}(S_i - O_i)}{n} \tag{2}$$

where $n$ is the total number of pixels, $O_i$ and $S_i$ are the 'observed' and 'simulated' values, respectively. When presenting the RMSE, the underlying assumption is that the errors are unbiased and follow a normal distribution (Chai and Draxler, 2014). In the realm of forecasting, Bias denotes a systematic deviation between the simulated data and the observed data, leading to an overestimation or underestimation.

In order to assess the precision of a binary segmentation analysis, such as flood inundation, two results corresponding to wet (i.e., flooded) and dry pixels are attainable (Peng et al., 2019). True Positive (TP), where wet pixels are accurately recognized as wet; True Negative (TN), where dry pixels are correctly classified as dry; False Positive (FP), where dry pixels are wrongly classified as wet, and False Negative (FN), wet pixels misclassified as dry pixels. Based on these outputs, pixel accuracy, which defines the proportion of properly categorized pixels, may be calculated. However, since accuracy computes this proportion regardless of classes, it might be deceptive when the class of interest (e.g., wet) contains a very small number of pixels. In order to circumvent this issue, the scores of Precision, Recall, and $F_1$ are often used. Precision is the percentage of correctly labeled wet pixels that were expected. However, Recall indicates how many pixels were correctly identified as wet by the ML-based predictive model. An accurate classification of an inundation map requires high levels of Precision and Recall. F1 score is often used as a tradeoff metric in this context to combine over- and under-segmentation into a single measure (Konapala et al., 2021).

$$F_1 = \frac{TP}{TP + \frac{1}{2}(FP + FN)} \tag{3}$$

In this study, instead of using actual field data, the outputs from the LISFLOOD-FP model are utilized as a baseline to evaluate the predictive abilities of the ML models with different spatial resolutions of DTMs and DSMs.

## 3. Study Area

The city of Carlisle, a medium-sized urban settlement in the United Kingdom, is used as the case study in this work to illustrate the effects of DEM resolution and model types on the performance of the CNN-based flood depth prediction model. Here, site-specific information as well as important hydrometric and spatial data are presented.

The city of Carlisle in Northwest England, which is located in the downstream Eden Catchment, is quite vulnerable to floods. Nearly 2500 km$^2$ of the Eden Catchment's drainage region gets an annual average precipitation of 1148 mm (Allen et al., 2010) with more than 150 wet days (i.e., with rainfall of ≥ 1 mm) (Met Office, 2023). The Carlisle region is predominantly characterized by its rural landscape, as urban land use comprises a mere two percent of the catchment area (Carlisle City Council, 2011). The research domain encompasses about 14.5 km$^2$ of Carlisle's urbanized region (Figure 2), which has historically experienced severe flooding, notably the floods of 2005 and 2015 (Liu et al., 2021). The study area encompasses the point at which three significant rivers, namely the Rivers Eden, Petteril, and Caldew, converge. As a result, fluvial flooding is the primary cause of flooding, accounting for 67% of all flooding incidents (Carlisle City Council, 2011).

The flooding that occurred in Carlisle in 2005 was selected for benchmarking and simulated using the LISFLOOD-FP. Heavy rains that fell across Northwest England on January 6-7, 2005, caused major flooding in and around Carlisle on January 8. Areas of the Eden watershed received up to 175 mm of rain in the 36 hours before the disaster (Day, 2005; Neal et al., 2009). The event began gradually, with flooding beginning early in the morning, well before the peak around midday. The majority of the residential/commercial sectors along the waterways and the low-lying rural regions towards the northeastern section of the city were flooded. It was calculated that the return period of this occurrence was about once every 150 years (Liu et al., 2021). Due to the large inundation extent of this event, there will be an opportunity to assess how well a model functions when subjected to various resolutions of DTM and DSM.

At the upstream sites of the three contributing rivers, boundary conditions are required to drive the flood event. The three boundary locations are shown in Figure 2 as "Upstream 1" in the northeast corner of the domain, "Upstream 2" and "Upstream 3" in the southern side. This study uses 15-minute discharge hydrographs for the 2005 event that were generated using 1D flood routing (Kabir et al., 2020). To train the ML model, 24 "synthetic" hydrographs—eight for each of the three upstream boundary locations—are created with variable peaks and durations to simulate alternative flood scenarios (Figure 3). For each upstream location, hydrographs corresponding to historical floods were chosen. As the peak discharge of these hydrographs was less than that of the January 2005 event, and the differences between the peaks were not significant enough for the study, the hydrographs were adjusted by multiplying them with a specific ratio of $Peak_{max}$ to $Q_{max}$. Here, $Peak_{max}$ represents the user-defined peak greater than the observed peak discharge, $Q_{max}$. This technique introduces greater variability in the peak discharge values of the input hydrographs. The data for the last three hydrographs (SH-6, SH-7, SH-8) comes from the past records kept at several gauging stations along the river Eden (Kabir et al., 2020). The hydrographs are generated or selected such that the flow in the river Eden is much bigger than the flows in the two tributaries to approximate reality.

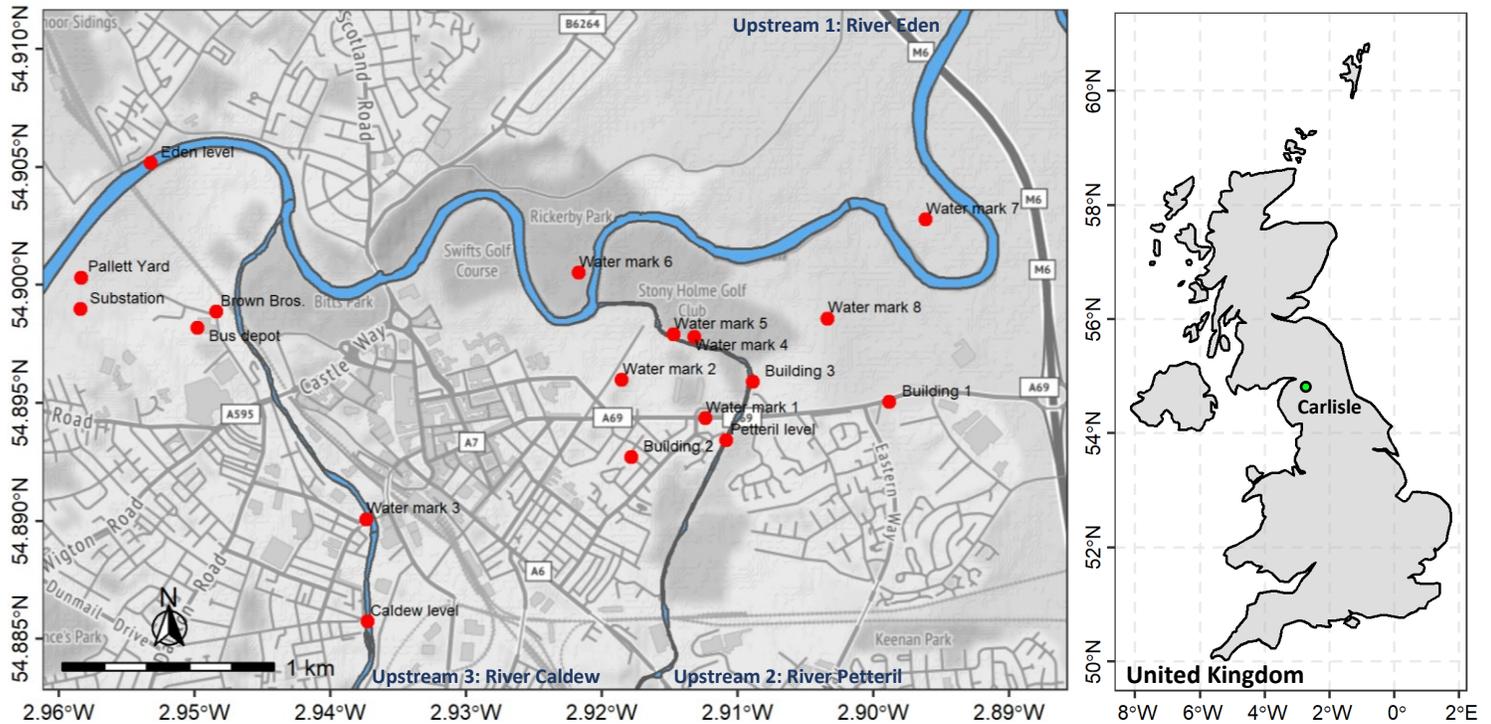

**Figure 2.** Study area as part of Carlisle City, UK, and 18 validation points within the domain.

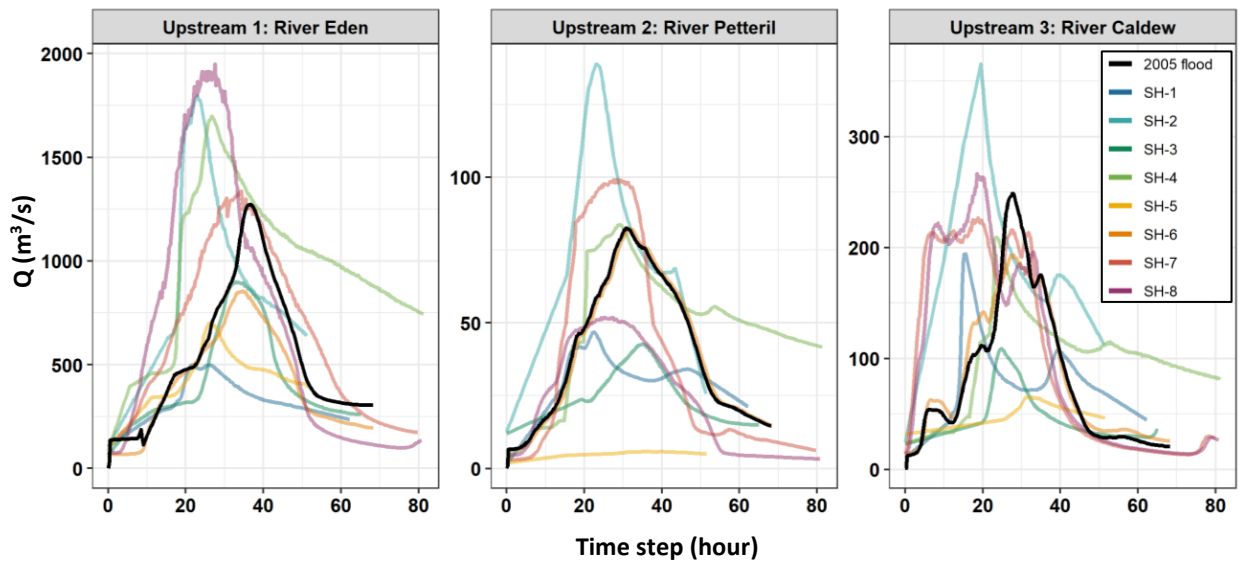

**Figure 3.** The 'synthetic' and routed/observed hydrographs used to train and test the ML models. For the 2005 Carlisle flood event, the hydrographs start at 00:00 h on 7th January 2005 as time 0. SH denotes synthetic hydrographs.

In this study, the LiDAR 1 m DTM released by the Environment Agency in 2020 with a vertical accuracy of +/-15cm RMSE (https://environment.data.gov.uk/) was further processed to obtain the DTM dataset of 5, 15, 20, 25, and 30 m resolutions. The associated DSMs were generated by combining the DTMs with the building footprint Polygons made available by OpenStreetMap (www.openstreetmap.org/). The DSM with a resolution of 5 m is considered to represent the baseline.

Different flood conditions at the study area characterized by DEMs are generated by running LISFLOOD-FP with eight sets of synthetic and one observed inflow hydrographs leading to training and test samples, respectively. The output files include 15-minute interval water depth grids in raster format. Each simulation has a uniform Manning coefficient $n$ of 0.055 sm$^{-1/3}$ across the whole domain. Here, a depth threshold of 0.3 m is implemented to exclude inconsequential depths from the target data (Kabir et al., 2020).

In this study, the key inputs for predicting water depths are discharge values with eight antecedent time steps and their corresponding observation time for each of the upstream locations. Therefore, timestamps, three upstream discharge values, and eight preceding discharges add up to a total of 28 input variables. The total number of samples amounts to 2104, which is the cumulative sum of the time steps from all eight synthetic hydrographs. Therefore, the input feature matrix takes the form of a 2104×28 matrix. The target matrix, on the other hand, depends on the resolution of each DEM and the total number of cells within the domain, as indicated in Table 2.

**Table 2.** Target matrix of the developed CNNs based on the resolution of DEMs.

| DEM Resolution (m) | Target matrix | | |
|---|---|---|---|
| | No. Rows (Train) | No. Rows (Test) | No. Columns (Train and Test) |
| 15 | 2104 | 266 | 64668 |
| 20 | | | 36414 |
| 25 | | | 23180 |
| 30 | | | 16218 |

The input feature matrix, consisting of 2104 samples, is constructed by vertically stacking the discharge with antecedent values and related observation periods for all synthetic hydrographs (see Figure 1). Moreover, flood depth raster data are read sequentially and then converted into arrays, which are then stacked vertically to form the target matrix. In order to get the most accurate results from a CNN-based model, it is necessary to optimize its hyperparameters. To fine-tune each of the prediction models, the Bayesian optimization approach is used. The data associated with the 2005 flood event is retained for use in model testing, while the eight synthetic flood scenarios are used for hyperparameter optimization across multiple DEMs.

Given that the synthetic hydrographs are fundamentally rooted in the watershed's actual response to hydro-meteorological conditions, the trained CNN-based model holds potential for real-world application. The model's ability to learn and generalize from these hydrographs can indeed be seen as a step toward applicability in practical scenarios. However, it is essential to emphasize that continuously updating and refining the input hydrographs used for training can improve the accuracy and adaptability of the CNN-based model. As the model learns from a broader and more diverse set of hydrographs that accurately represent evolving real-world conditions, its performance and accuracy in predicting and responding to flood events can be enhanced.

# 4.   Results and Discussion

This section presents a comprehensive analysis of the flood inundation maps, focusing on two key comparisons. Firstly, the variation of flood inundation depth is compared with a baseline reference. Secondly, the flood inundation extent is examined, enabling the identification of spatial patterns, trends, and potential impacts. By delving into the findings and engaging in a thoughtful discussion, a deeper comprehension of the flood event and its consequences can be achieved.

## 4.1   Comparison of different resolutions in selected stations

The flood inundation map depicted in Figure 4 was generated using LISFLOOD-FP and a DSM with a resolution of 5 m. The inundation maps delineate four distinct phases: early (Figure 4a), growth (Figure 4b), peak (Figure 4c), and recession (Figure 4d), each occurring at 12-hour intervals. Notably, observations indicate that the flood peak at Upstream 1 was noted around 12:00 on January 8, 2005, approximately 36 hours after the onset of the flood. In contrast, for the smaller rivers (Upstream 2 and 3), the flood peak occurs slightly earlier. However, the magnitude of peak flooding in these smaller upstream areas (83 and 249 m$^3$/s) is notably lower compared to Upstream 1 (1272 m$^3$/s). Consequently, it can be inferred that the peak inundation in the study area aligns with the timing observed in the main and larger stream (Upstream 1). The corresponding inundation area values for these four stages are 1.86, 4.02, 5.11, and 4.41 km$^2$, respectively.

Figure 5 presents the CNN-based flood depth variations for different DEM resolutions and types in some selected locations such as the Bus depot, Caldew level, Watermark 3, and Eden level stations. In the Bus depot, the water depth is increased as the resolution of the DTM becomes coarser, and it has no effect on the DSM lower than 30 m resolution. The location of station near the riverbanks increases the likelihood of being directly influenced by rising water levels and potential overflow. The surrounding topography, characterized by low-lying areas or inadequate drainage systems, further contributes to the station's vulnerability. In the Caldew level which is located at the upstream section of the river Caldew, there is a noticeable abrupt increase in water depth during the peak flood period in both DTM and DSM with 25 m resolution. However, it is important to note that the water depth remains relatively consistent around the peak time, ranging between 4.5 to 7.5 m. Watermark 3, which is situated on the lower side of the Caldew level station, close to the river Caldew, has less than 4.5 m water depth for the DSM and DTM of lower than 30 m resolution. During the flood event, it encountered a minimum water depth of at least 3 m at a resolution of 30 m, with the maximum water depth occurring at the peak time. The differences are most evident at the Eden level station, which experiences greater inundation depth compared to the other stations. It might be due to the fact that the station is located at the end of the main river and receives the highest volume of water, particularly during peak times and in coarse resolutions. At the Eden level station, the flood depths recorded in both DSM and DTM remain consistently similar across various resolutions. This can be attributed to the station's proximity to the river. In this case, the absence of buildings near the river means that the DSM, which accounts for the elevation of structures, does not significantly impact the water depths. Consequently, the flood depths obtained from both the DSM and DTM provide comparable results. This proximity to the river and the absence of structures influencing the water levels contribute to the reliability and consistency of the flood depth measurements at the Eden level station. The inundation depth variation for the rest of the validation points can be found in Figures S1 to S14 in the supplementary information (SI).

The findings demonstrated a consistent trend of increasing water depth in most stations as the resolution of the DEMs became coarser. The correlation between coarser DEM resolutions and higher water depths in flood modeling can be attributed to several factors. Coarser resolutions are linked to higher water depths because of their difficulty in capturing small-scale terrain features. High-resolution DEMs catch details like embankments and slopes that significantly influence water flow during floods. As the resolution coarsens, it struggles to depict these details accurately. Coarser resolutions tend to smooth out elevation differences, especially in areas like densely built cities (Zheng et al., 2018; Xu et al., 2021). In addition, coarser resolutions might oversimplify channel networks, affecting water routing and contributing to the association with higher water depths (Hou et al., 2021). Once the DEM resolution exceeds the river width, the overprediction of flood depth and extent increases significantly (Muthusamy et al., 2021). However, Fatdillah et al. (2022) observed that in certain situations, the opposite effect can occur, where a finer resolution of the DEM leads to an expansion of the inundation area.

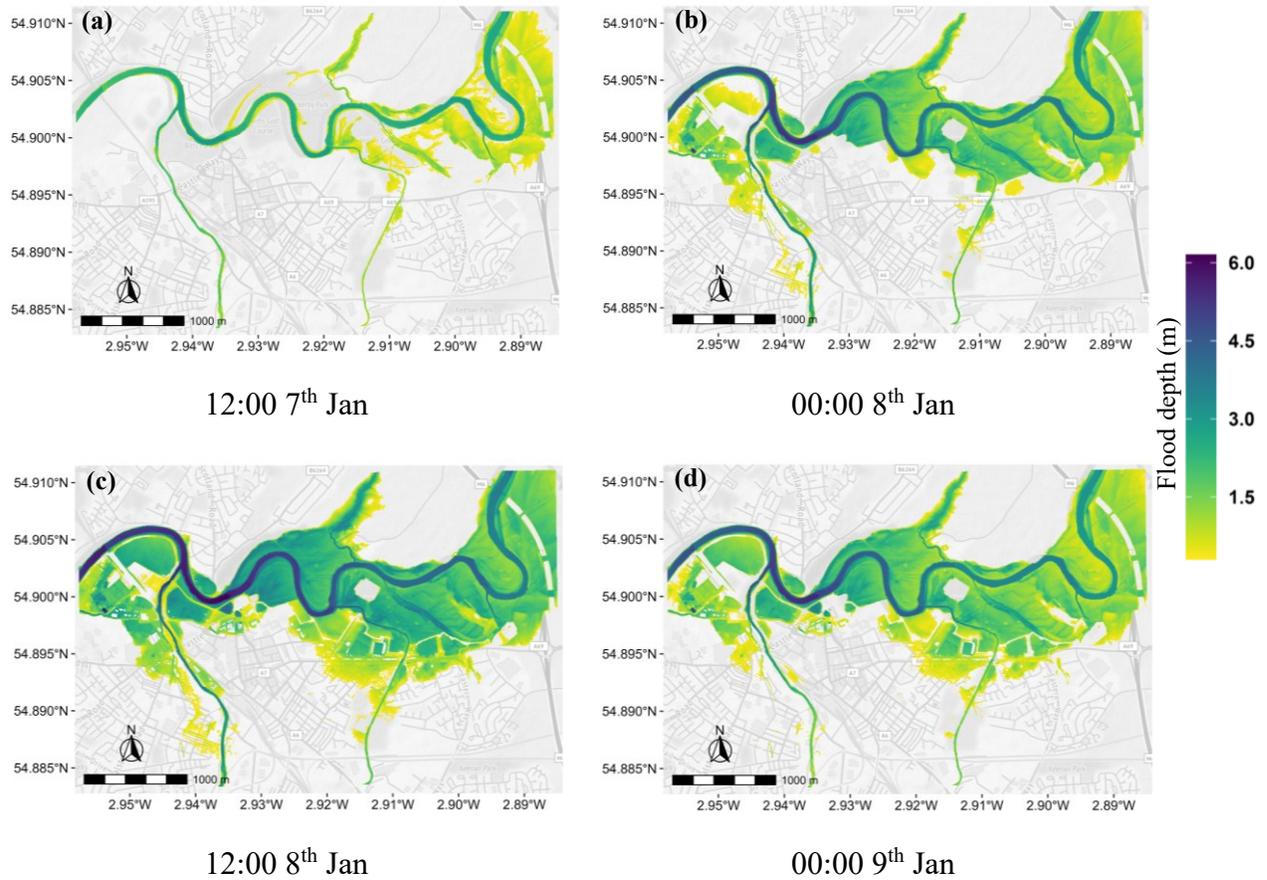

**Figure 4.** Flood maps generated by LISFLOOD-FP for 5 m resolution LiDAR DSM in four timestamps, all referencing the 2005 event and serving as the baseline.

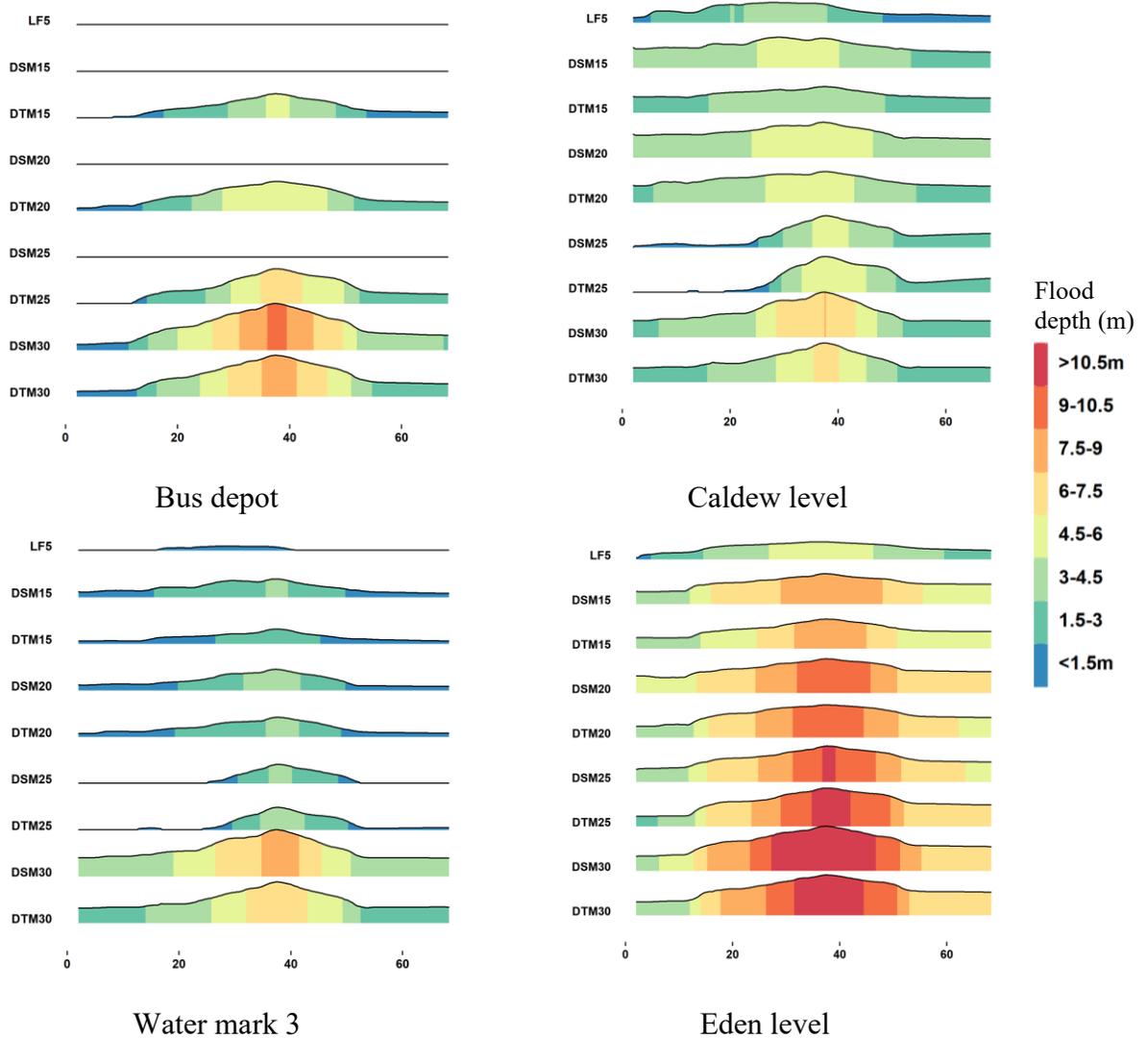

**Figure 5.** Water depths of Four stations for different resolutions from obtained CNN and LISFLOOD-FP. The x-axis (hour) shows the time series of the flood.

## 4.2 Inundation error maps

To further examine the robustness of the models, spatial error maps were produced and presented in Figure 6. The left column shows the result for DTM15, and the right column shows the result for DSM15, illustrating the absolute difference in water depth between the CNN-based and LISFLOOD-FP inundation maps using a 15 m resolution for the 2005 event. In the error maps, it is evident that the DTM exhibits lower error rates compared to the DSM in regions devoid of buildings, with errors primarily concentrated at the peak of the flood event. During the initial stage of the flood, the DSM map shows the highest errors in the vicinity of the main river, specifically in the areas surrounding the Brown Bros and Bus depot stations, extending towards the Swifts Golf Course. Notably, since this region lacks buildings, the application of DSM has no discernible impact on improving flood prediction by CNN. These errors gradually diminish as the flood progresses and become prominent in certain grassland areas, as depicted in Figure 6d. As the flood intensifies, the DSM map experiences maximum errors during the peak flood, in contrast to the DTM. Additionally, there are two small green areas surrounding Building 1 and Building 2 stations in the DSM maps (Figure 6f and Figure 6h) that exhibit the lowest error rates. This indicates that the implementation of DSM positively influences flood inundation mapping for these built-up areas.

### 4.3 Inundation error indices

To make a quantitative comparison between the machine learning (ML) predictions and the reference results obtained from LISFLOOD-FP, various error measures such as RMSE (Root Mean Square Error) and Bias are computed at 18 control sites. This analysis is conducted for both DTMs and DSMs with spatial resolutions of 15, 20, 25, and 30 m. When considering RMSE and Bias, the performance of DTMs is slightly superior to that of DSMs across all resolutions during the flood initiation (48[th] stage). During this stage, the RMSE and Bias values for the DTM15 indicated an inundation depth error of approximately 0.65 and 0.45 m, respectively. These values appear to fall within an acceptable range. The positive sign of the Bias suggests that the predicted inundation depth was overestimated. In contrast, the performance of DSMs and DTMs of 15 m and 20 m resolution appears to be more comparable and similar during the remaining stages of the flood. Additionally, DTM30 shows lower RMSE and Bias values in comparison to the DSM30, implying that the DTM provides a slightly more accurate estimation of flood conditions. It is important to highlight that the RMSE values exhibit significant increases as the time steps progress. Specifically, there is an approximately 2-fold increase from the 48th to 96th time steps, a 4-fold increase from the 48th to 144th time steps, and a slightly less than 3-fold increase from the 48th to 192nd time steps. These findings suggest that the accuracy of the predictions deteriorates as the flood event progresses, indicating the potential limitations of the model over longer durations of large floods. Furthermore, the Bias values demonstrate a substantial increase starting from the early stages of the flood, reaching their peak at the time of maximum inundation. This significant increase in Bias indicates a consistent tendency of the model to overestimate the flood depth as the event unfolds.

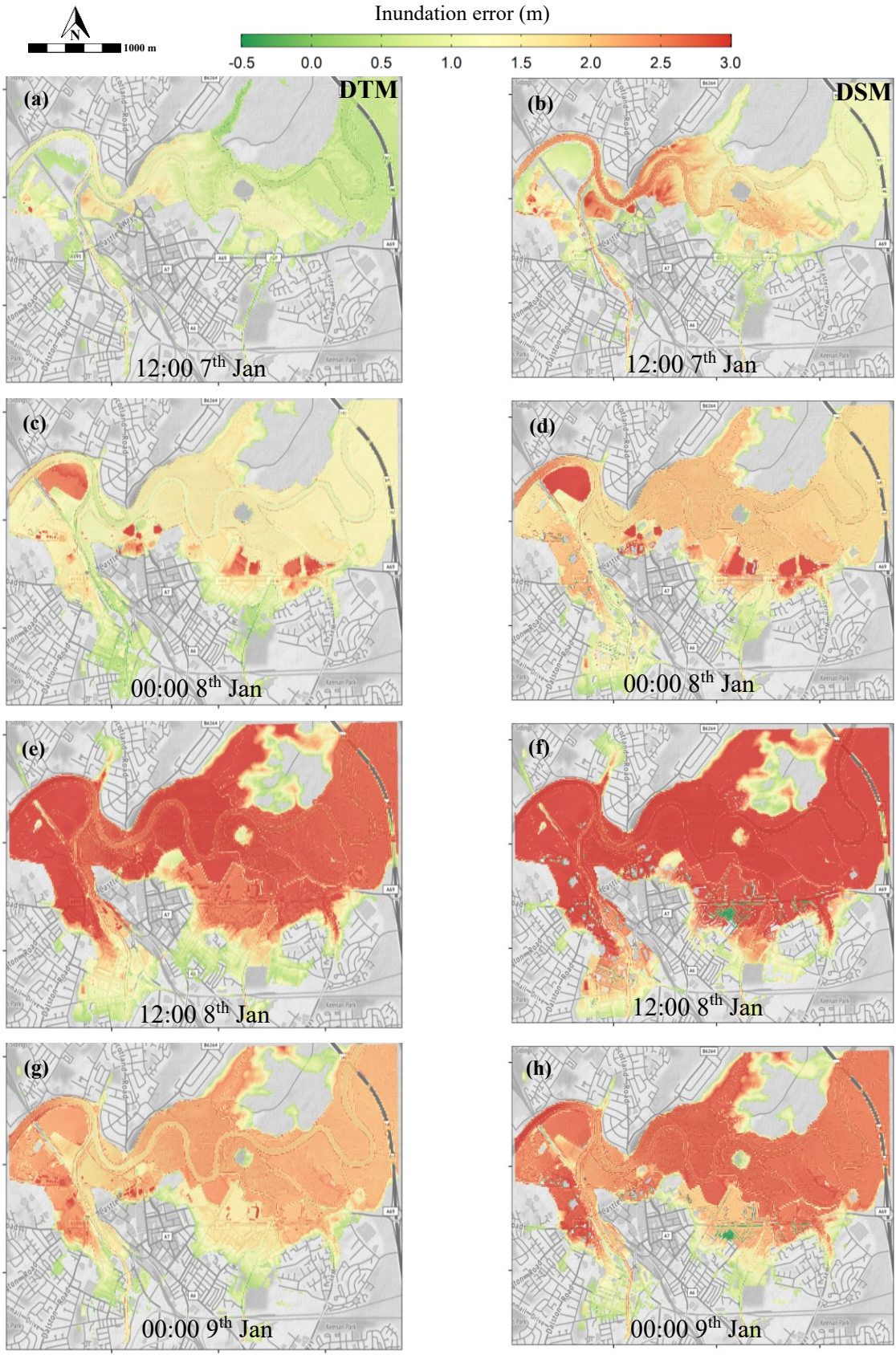

**Figure 6.** Error maps at four selected time stamps for DTM15 and DSM15.

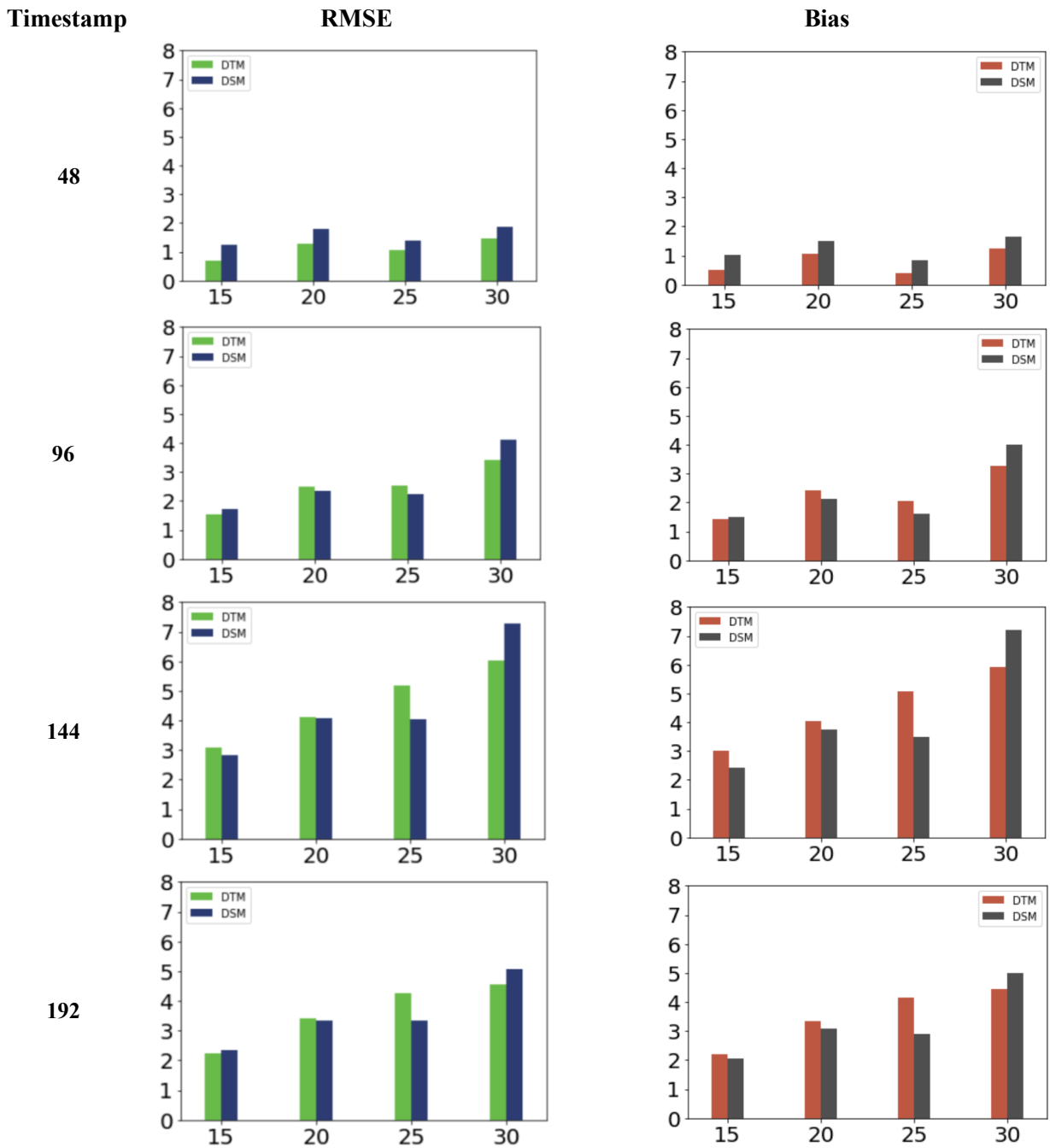

**Figure 7.** Average of RMSE and Bias by the CNN-2005 model in the 18 control locations for DTM and DSM in four stages for different four resolutions.

The performance of the CNN-based models using DTM and DSM of different resolutions in predicting the inundation extent is presented in Table 3, assessing precision, recall, and F1 metrics. Precision indicates the proportion of correctly identified inundated cells in the predicted map out of the total number of predicted inundated cells. Recall measures the ratio of correctly identified inundated cells to the sum of true inundated cells and falsely classified non-inundated cells in the predicted map. Lastly, the F1 metric serves as a composite index that combines precision and recall providing an overall evaluation of accuracy.

During the early stage of the flood (48th time step), the accuracy of DTM values is notably higher, particularly in finer resolutions. For instance, at resolutions of 15 and 20 m, the F1 scores are 0.644 and 0.538 for DTM, respectively, compared to 0.573 and 0.522 for DSM. This discrepancy can be attributed to the limited extent of inundation at the onset of the flood, primarily affecting bare areas near the river. However, as time passes and the flood expands, more buildings come under the influence of the flood, resulting in DSM demonstrating greater accuracy in comparison to early stages and DTMs of coarser resolutions. This trend is more evident at the peak flood (144th time step), where DSM exhibits greater precision and F1 scores for all selected spatial resolutions. The same holds true for DSM of resolution higher than 20 m for the second and last stages (i.e., 96th and 192nd time steps). Moreover, the recall values approach a high value of 1, indicating accurate identification of true inundated pixels. However, this accuracy is achieved at the expense of overestimating the flood predictions based on coarser resolution DEMs (Ogania et al. 2019; Karamouz and Fereshtehpour 2019). It is important to note that the F1 score, which combines precision and recall into a harmonic mean, places greater emphasis on smaller values. Upon examination of Table 3, it is evident that the recall values remain consistently high across various resolutions and stages, albeit with lower precision values. As a result, the F1 performance aligns more closely with precision.

It is worth noting that factors such as the resolution of the data, the availability and quality of elevation data, and the accuracy of flood prediction models can also influence the accuracy of DTM and DSM values throughout the flood event. Furthermore, the effectiveness of DTM and DSM may vary depending on the specific geographical and environmental characteristics of the area being studied. The DSM is typically employed in densely built areas to evaluate the impact of buildings on flood-prone regions, and it is most effective when the buildings cover the entire cell. In this study, however, the selected spatial resolutions are relatively low, and the size of each cell ranges from 225 m$^2$ (15m×15m) to 900 m$^2$ (30m×30m). Additionally, the cells might not be fully occupied by existing buildings in reality, making it difficult for the model to produce accurate estimates. Consequently, for the areas with a low density of buildings in the early stages of the flood where the inundation is low, DTM has better performance in these low resolutions. However, in the next stages, including growing and peak, DSM is more accurate. This is in line with Shen and Tan et al. (2020) who compared various modeling scenarios with different DEM resolutions to investigate DEM sensitivity to the building treatment method (BTM) and urban inundation modeling. Because of the obstruction of flow routes between building elements, modeling scenarios that are coarser than the size of the buildings and gaps lead to poor performance in inundation simulations.

**Table 3.** Spatial accuracy scores of the CNN-2005 versus LISFLOOD-FP during the flood initiation, growing, peak, and recession stages for DTM and DSM.

| Index | DEM Resolution (m) | | Time step | | | |
|---|---|---|---|---|---|---|
| | | | 48 | 96 | 144 | 192 |
| Precision | DTM | 15 | 0.476 | 0.652 | 0.624 | 0.662 |
| | | 20 | 0.368 | 0.563 | 0.550 | 0.559 |
| | | 25 | 0.425 | 0.601 | 0.528 | 0.492 |
| | | 30 | 0.315 | 0.470 | 0.467 | 0.481 |
| | DSM | 15 | 0.402 | 0.641 | 0.645 | 0.652 |
| | | 20 | 0.354 | 0.606 | 0.587 | 0.574 |
| | | 25 | 0.381 | 0.646 | 0.596 | 0.582 |
| | | 30 | 0.316 | 0.481 | 0.497 | 0.496 |
| Recall | DTM | 15 | 0.992 | 0.997 | 1.000 | 0.999 |
| | | 20 | 0.995 | 0.999 | 1.000 | 1.000 |
| | | 25 | 0.972 | 0.983 | 0.998 | 0.998 |
| | | 30 | 0.986 | 0.994 | 0.996 | 0.995 |
| | DSM | 15 | 0.996 | 0.995 | 0.981 | 0.986 |
| | | 20 | 0.996 | 0.994 | 0.980 | 0.985 |
| | | 25 | 0.982 | 0.977 | 0.976 | 0.982 |
| | | 30 | 0.988 | 0.988 | 0.973 | 0.978 |
| F1 | DTM | 15 | 0.644 | 0.789 | 0.768 | 0.797 |
| | | 20 | 0.538 | 0.720 | 0.710 | 0.717 |
| | | 25 | 0.591 | 0.746 | 0.691 | 0.659 |
| | | 30 | 0.477 | 0.638 | 0.636 | 0.648 |
| | DSM | 15 | 0.573 | 0.780 | 0.778 | 0.785 |
| | | 20 | 0.522 | 0.753 | 0.734 | 0.726 |
| | | 25 | 0.549 | 0.778 | 0.740 | 0.731 |
| | | 30 | 0.479 | 0.647 | 0.658 | 0.659 |

#### 4.4 Applicability in a data-scarce region

Building on the insights gained from the study conducted in Carlisle, UK, we now discuss the applicability of the developed deep learning models to address the challenges of data-scarce flood-prone regions, exemplified by Pakistan. This country has been increasingly vulnerable to frequent and severe floods, causing extensive damage (Hashemi et al. 2012; Manzoor et al. 2022). Given this pressing challenge, recent research endeavors have concentrated on harnessing the power of machine learning to predict and map flood inundation (Yaseen et al. 2022a and b). Recently, a study employed a combination of remote and social sensing techniques, along with geospatial data and advanced machine learning approaches, to comprehensively map flood exposure, assess damage, and address population needs during the 2022 Pakistan floods (Akhtar et al. 2023). In another study focused on the Jhelum River in the Punjab region, Ahmad et al. (2022) implemented various artificial intelligence techniques such as local linear regression, dynamic local linear regression, and artificial neural networks to enhance flood prediction accuracy. Additionally, innovative machine learning models driven by rainfall data were successfully employed to identify flood-prone areas in Karachi, Pakistan, while an ensemble machine learning approach proved effective in mapping flood susceptibility in an arid region of the country (Yaseen et al. 2022; Rasool et al. 2023). These endeavors underscore the substantial potential of machine learning in predicting and mapping flood inundation in Pakistan, thus offering invaluable tools for bolstering disaster management and mitigation strategies. However, the use of digital elevation model (DEM) data in flood inundation mapping is limited by its low spatial resolution, which constrains the accuracy and applicability of ML methods.

The availability of DEM data in Pakistan can vary throughout the country. Ahmad et al. (2010) noted that the highest available DEM resolution in Pakistan is 30 m. However, numerous studies have successfully utilized this data for various purposes. For example, 30 m DEM data obtained from the Japan Aerospace Exploration Agency (JAXA) was processed in ArcGIS to identify potential runoff harvesting sites in the Karoonjhar Mountainous Area (Siyal et al., 2018). Moreover, in recent flood mapping and assessment studies on 2022 mega flood conducted in Pakistan, the 30 m STRM DEM data has been employed in conjunction with satellite imagery (Wang et al., 2023). These examples demonstrate the utilization of DEM data in different research applications in Pakistan. Efforts should continue to ensure the availability and accessibility of high-quality DEM data across all country regions to support accurate flood mapping and prediction (Ahmad et al., 2010).

Given DEMs with a spatial resolution of 30 m, it is anticipated that the use of digital surface models instead of digital terrain models will result in reduced accuracy throughout all flood stages. When considering the specific case of Carlisle, utilizing a 30 m DSM is anticipated to result in an increase in RMSE compared to using DTMs, across all flood stages. This increase in RMSE is estimated to be around 30% during the flood initiation stage, 21% during the peak stage, and 12% during the recession stage. Furthermore, a similar range of reduction is observed for the Bias, indicating a tendency to overestimate the results when employing DSMs instead of DTMs. In terms of the overlap score (F1), the use of DSMs demonstrates a slight improvement compared to DTMs, with the peak stage exhibiting the highest increase of 3.5%. However, this analysis reveals that DTMs generally outperform DSMs. This suggests that considering the current DEM availability, DTMs are the recommended choice for flood mapping applications.

From the spatial resolution lens, employing a 15 m DTM leads to a significant increase in the accuracy of flood inundation depth, with RMSE and bias improving by approximately 50% compared to the 30 m DTM during all phases of the flood. In terms of the overlap index, elevating the DTM resolution from 30 m to 15 m results in a minimum 20% enhancement in the accuracy of flood inundation extent, with the flood initiation stage exhibiting the highest improvement of 35% compared to the 30 m DTM. When comparing a 15 m DSM to a 30 m DTM, the F1 score shows a consistent improvement across all flood phases, increasing by approximately 22%. Furthermore, taking into account the F1 score of a higher resolution 5 m DSM, as reported by Kabir et al. (2020), boosts this score by over 50%, notably by 55% during the peak stage compared to a 30 m DTM.

In addition to accuracy considerations, it is crucial to account for runtime when evaluating the overall efficiency of flood prediction models. In general, while higher resolution DEMs can enhance the accuracy of flood mapping, they also come with increased computational demands. The 1D CNN model used in this study is optimized for acceleration through Graphics Processing Units (GPUs). The CNN models developed for all DTMs and DSMs were trained on an NVIDIA GeForce 940MX. At a spatial resolution of 15 m, the domain contains 64,668 cells, while at a 30 m resolution, there are 23,180 cells. The corresponding runtimes are approximately 6 and 3 minutes, respectively. According to Kabir et al. (2020), employing a more advanced GPU dedicated to workstations (Tesla P100 GPU) for a 5 m resolution DSM with 581,061 cells in the domain resulted in a total training and testing time of 5 minutes (which is tens of times faster than simulating using hydrodynamic models). Hence, for the small domain size in this study, the overall runtime has a minor impact on the 1D CNN model, and the model's efficiency is primarily influenced by its accuracy. However, for larger domains like an entire watershed, the computational workload could significantly affect the overall efficiency of the model.

These compelling results indicate the added value of higher resolution DEMs and highlight the significance of allocating resources toward data acquisition to facilitate advanced flood mapping techniques. The inadequacy of the current DEM resolution in Pakistan underscores the need for higher resolution data to achieve precise and reliable flood inundation mapping. Even a 15 m DTM can have a significant impact on flood inundation assessments (Fereshtehpour and Karamouz, 2018), thereby enhancing risk and disaster management efforts. Therefore, continued efforts are necessary to improve the availability and accessibility of high-quality, high-resolution DEM across all regions, maximizing the potential of advancing machine learning techniques.

## 5. Conclusion

This study delved into the intricate relationship between Digital Elevation Models, deep learning techniques, and flood inundation modeling. Despite the growing interest in utilizing machine learning approaches for flood prediction and risk assessment, this specific domain has received limited attention. To fill this gap, we initially compiled a comprehensive summary of existing literature on flood prediction using Machine Learning, aiming to consolidate knowledge in the field. Subsequently, our study focused on evaluating the effectiveness of a CNN-based flood prediction model in an urban setting by investigating different resolutions of DTMs and DSMs. Through the utilization of a hydrodynamic LISFLOOD-FP model and LIDAR-based 1 m resolution DTM, the study analyzed fluvial flooding in the city of Carlisle, UK. In densely built-up areas with high spatial variability, our study revealed that the sensitivity of simulated inundation levels to DEM resolution and resampling approaches is particularly pronounced. We found that, within the range of resolutions examined, using a DTM yields slightly lower RMSE and reduced bias compared to a DSM. In the case of Carlisle, employing a 30 m DSM resulted in a 21% increase in RMSE during the peak stage compared to the 30 m DTM. In addition, employing a 15 m resolution yields a superior accuracy in inundation extent prediction compared to 30 m DTM, with an average enhancement of 26% and 21.5% for DTM and DSM, respectively, across all flood phases. These findings underscore the importance of improving the availability and accessibility of high-resolution DEMs, particularly in data-scarce regions like Pakistan, as they can significantly impact risk management efforts.

While higher-resolution DEMs generally offer more accurate results for flood inundation mapping, a careful balance between resolution, computational resources, and processing time is required when choosing the suitable DEM resolution. 1D CNNs are known for their memory inefficiency, limiting their applicability to smaller domains. Therefore, in future research endeavors involving higher spatial resolutions, it becomes necessary to either reduce the domain size or explore alternative network architectures (e.g., 2D convolutions) that can handle the increased memory demands.

# Declarations

**Funding** This research was partly supported, with funding provided to the third author, by the National Oceanic and Atmospheric Administration (NOAA) through an award to the Cooperative Institute for Research to Operations in Hydrology (CIROH) (NOAA Cooperative Agreement with The University of Alabama, NA22NWS4320003).

**Authors' contributions** *Conceptualization*: Mohammad Fereshtehpour; *Data Curation*: Mostafa Esmaeilzadeh; *Methodology*: Mohammad Fereshtehpour; *Formal analysis and investigation*: Mohammad Fereshtehpour, Reza Saleh Alipour, Mostafa Esmaeilzadeh; *Writing - original draft preparation*: Mohammad Fereshtehpour, Mostafa Esmaeilzadeh, Reza Saleh Alipour;

*Visualization*: Mohammad Fereshtehpour, Reza Saleh Alipour; *Writing - review and editing*: Steven J. Burian, Mohammad Fereshtehpour; *Funding acquisition*: Steven J. Burian.

**Acknowledgments** We would like to thank Sayed Rezwan Kabir and Mehdi Aboosaeedi for their invaluable assistance during the early stages of this research and for providing essential input data.

**Data availability** The datasets generated during and/or analyzed during the current study are available from the corresponding author on request.

**Conflict of interest** The authors declare that they have no known competing financial interests or personal relationships that could have appeared to influence the work reported in this paper.